\documentclass[a4paper,11pt]{article}
\usepackage{pos}
\usepackage{float}
\usepackage{siunitx}
\usepackage{booktabs}
\usepackage{mathtools}
\usepackage{bbm}
\usepackage[noabbrev, nameinlink]{cleveref}
\usepackage{layouts}
\title{Comparison of topology changing update algorithms}

\author*{Timo Eichhorn}
\author{Christian Hoelbling}

\affiliation{Department of Physics, University of Wuppertal, Gaußstraße 20, D-42119 Wuppertal, Germany}

\emailAdd{timo.eichhorn@protonmail.com}
\emailAdd{hch@uni-wuppertal.de}

\abstract{In modern lattice simulations, conventional update algorithms do not allow for tunneling between topological sectors at fine lattice spacings. We compare the viability of multiple less commonly used algorithms (metadynamics, instanton updates, and multiscale thermalization) with respect to proper sampling of all topological sectors in the Schwinger model. We briefly comment on the prospects of applying these methods to 4-dimensional SU(3) simulations.}

\FullConference{%
 The 38th International Symposium on Lattice Field Theory, LATTICE2021
  26th-30th July, 2021
  Zoom/Gather@Massachusetts Institute of Technology
}

\begin{document}
\maketitle
\newcommand\thefont{\expandafter\string\the\font}

\section{Introduction}
An important property that has to be taken into consideration when analyzing data generated by Markov chain Monte Carlo methods is the correlation of subsequent states in the Markov chain. A particularly problematic form of these autocorrelations in the context of physical simulations is critical slowing down: Near critical points, many autocorrelation times associated with observables measured on the generated configurations diverge.
\begin{figure}[H]
    \centering
    $\vcenter{\hbox{\includegraphics[width=0.495\textwidth]{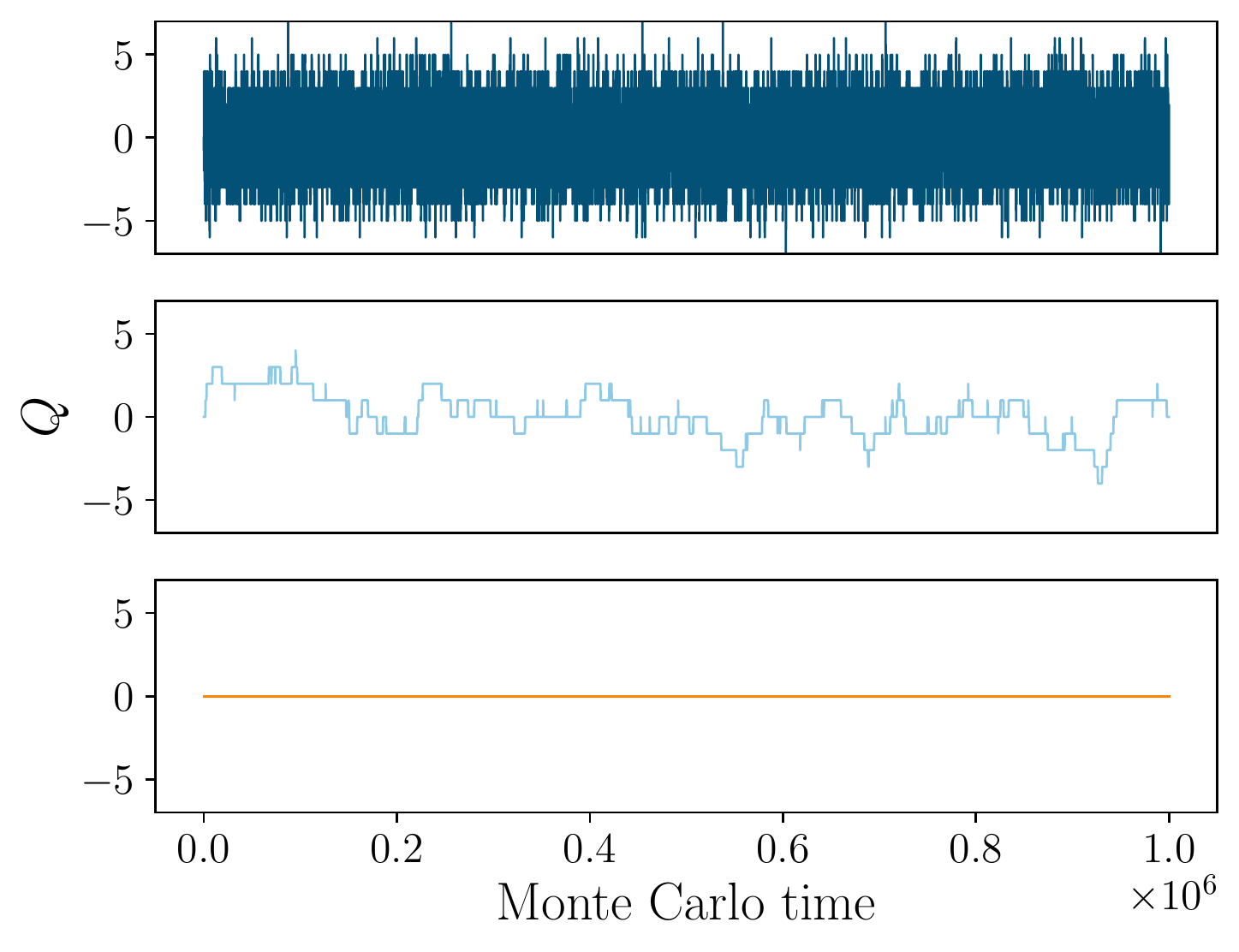}}}$
    $\vcenter{\hbox{\includegraphics[width=0.495\textwidth]{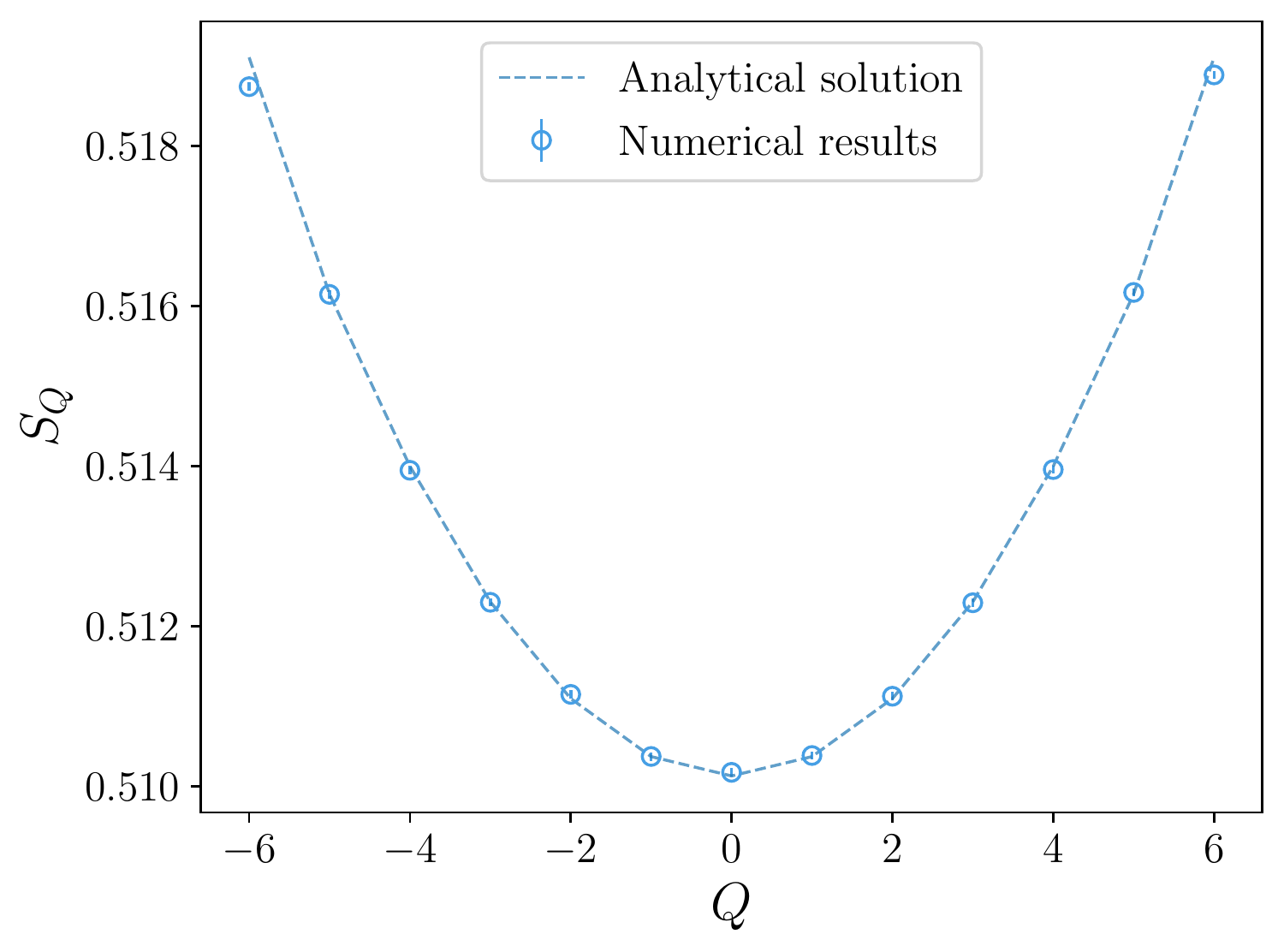}}}$
    \caption{\textbf{Left:} Time series of the topological charge in the Schwinger model for three simulations on a line of constant physics with the parameters $V = \{16^{2}, 24^{2}, 32^{2}\}$ and $\beta = \{3.2, 7.2, 12.8\}$ (from top to bottom). \textbf{Right:} Expectation values of the gauge action in the Schwinger model in different topological sectors for a $32^{2}$ lattice at $\beta = 12.8$ obtained through simulation and an analytical solution \cite{Kovacs:1995nn, Elser:2001pe, Bonati:2019ylr, Bonati:2019olo}.}
    \label{fig:figure1}
\end{figure}
In lattice QCD and lattice gauge theory in general, one prominent example of critical slowing down is related to the topological charge when taking the continuum limit \cite{Alles:1996vn, DelDebbio:2002xa, Schaefer:2009xx}. The autocorrelation time associated to the topological charge increases with the inverse lattice spacing, and after a certain point, the Markov chain may even be completely frozen in a topological sector for all practically achievable simulation timescales. The left plot in \cref{fig:figure1} illustrates this phenomenon in the Schwinger model for three different lattice spacings. Whereas transitions between different topological sectors happen frequently on the coarsest lattice, the Markov chain is confined to the $Q = 0$ sector throughout the entire simulation for the finest lattice spacing. This not only affects the estimation of statistical errors, but also has direct consequences on the expectation values of observables - even those not directly related to the topological charge. For instance, the average action increases for sectors with a larger absolute value of the topological charge, as shown in the right plot in \cref{fig:figure1}. Therefore, it is more accurate to describe the time evolution of the Markov chain in terms of modes that couple differently to observables, instead of simply associating a single autocorrelation time to each observable. In this case, the topological charge is strongly coupled to slow modes, whereas the action is not coupled as strongly to those modes.

While topological freezing appears in both the Schwinger model and four-dimensional SU(3) (with and without dynamical fermions), the results presented here are mostly limited to the former. We study three previously suggested algorithmic approaches to alleviate the problem: Metadynamics \cite{Laio_2002, Laio:2015era}, instanton updates \cite{Fucito:1983qg, Dilger:1994ma, Durr:2012te}, and multiscale thermalization \cite{Endres:2015yca, Detmold:2016rnh, Detmold:2018zgk}. 

\section{Metadynamics}
Metadynamics is based upon introducing a set of collective variables (CVs) to measure which parts of the configuration space have already been visited by the Markov chain. The space spanned by the CVs can thus be understood as a low-dimensional projection of the complete configuration space. Using the CVs, a history-dependent metapotential $V_{\mathrm{meta}}$ is built up, which has been shown to be an estimate of the negative free energy/action of the system \cite{PhysRevLett.96.090601}. During each accept-reject step of the update algorithm, the metapotential is added to the action of the configuration. Therefore, once the metapotential has reached equilibrium, the probability distribution will be approximately constant as a function of the CVs.

Following the approach of \cite{Laio:2015era} we use a linearly interpolating histogram instead of the originally proposed sum of Gaussians, which allows us to keep the memory footprint constant in time. The metapotential is defined by four parameters: $Q_{\mathrm{min}}$ and $Q_{\mathrm{max}}$ specify the interval $[Q_{\mathrm{min}}, Q_{\mathrm{max}})$ in which the metapotential is updated, $\delta Q$ is the bin width, and $w$ is a weight that controls how fast the metapotential is updated. As CV we use the imaginary part of all plaquettes, which yields a non-integer definition of the topological charge:
\begin{equation}
    Q_{\mathrm{meta}} = \frac{1}{2 \pi} \operatorname{Im}\left(\sum\limits_{\mathbf{n} \in V(\Lambda)} P_{tx}(\mathbf{n})\right).
\end{equation}
$P_{tx}(\mathbf{n})$ denotes the plaquette in $t$-$x$-direction at the point $\mathbf{n}$, and $V(\Lambda)$ refers to the set of all lattice points. At Monte Carlo time $t$ with a configuration of topological charge $Q_{\mathrm{meta}}(t)$, the metapotential is updated in the following way:
\begin{equation}
\begin{aligned}
    V_{\mathrm{meta}, i}(t) &= V_{\mathrm{meta}, i}(t - 1) + w \left( 1 - \frac{Q_{\mathrm{meta}}(t) - (Q_{\mathrm{min}} + i \cdot \delta Q )}{\delta Q} \right),
    \\V_{\mathrm{meta}, i + 1}(t) &= V_{\mathrm{meta}, i + 1}(t - 1) + w \frac{Q_{\mathrm{meta}}(t) - (Q_{\mathrm{min}} + i \cdot \delta Q )}{\delta Q}.
\end{aligned}
\end{equation}
The index $i$ is defined as
\begin{equation}
    i = \lfloor \frac{Q_{\mathrm{meta}}(t) - Q_{\mathrm{min}}}{\delta Q} \rfloor.
\end{equation}
In the end, the observables need to be reweighted back to the original probability distribution:
\begin{equation}
    \langle O \rangle = \frac{\sum\limits_{t} O(t) e^{\Bar{V}_{\mathrm{meta}}(Q_{\mathrm{meta}}(t))}}{\sum\limits_{t} e^{\Bar{V}_{\mathrm{meta}}(Q_{\mathrm{meta}}(t))}}.
\end{equation}
Here, $\Bar{V}_{\mathrm{meta}}$ refers to an estimate of the negative action, which can for instance be obtained via the time average of the metapotential (after excluding equilibration) or simply $V_{\mathrm{meta}}(t)$ for large $t$.

In practice, it is advisable to introduce an additional penalty potential to prevent the Markov chain from getting stuck in regions outside the interval $(Q_{\mathrm{min}}, Q_{\mathrm{max}})$ where the metapotential is not updated. The exact form of the penalty potential is not important - in our case we used a penalty potential that scales quadratically with the distance and depends on three parameters $k$, $Q_{\mathrm{thr,\, min}}$, and $Q_{\mathrm{thr,\, max}}$:
\begin{equation}
    V_{\mathrm{pen}} =
    \begin{cases}
    k \cdot \min\{(Q - Q_{\mathrm{thr,\, min}})^{2}, (Q - Q_{\mathrm{thr,\, max}})^{2}\} & Q \notin ( Q_{\mathrm{thr,\, min}},  Q_{\mathrm{thr,\, max}}),
    \\ 0 & \mathrm{otherwise.}
    \end{cases}
\end{equation}

\section{Instanton updates}
Instanton updates, which resemble another recently proposed update algorithm \cite{Albandea:2021lvl}, generate new configurations by multiplying the previous configuration with an instanton of charge $Q_{j} = \pm 1$, where the sign is chosen randomly with equal probability (for recent applications see \cite{Durr:2021gma, Frech}). In $D < 4$ dimensions, instantons are maximally delocalized, i.e., the field strength is constant over the entire lattice. In the temporal gauge on a $N_{x} \times N_{t}$ lattice with periodic boundaries, instantons of charge $Q_{j}$ can be constructed by setting the link variables to \cite{Smit:1986fn}:
\begin{equation}
\begin{aligned}
    U_{t}^{I}(Q_{j}; t, x) &= \exp\left( -2 \pi i x \frac{ Q_{j}}{N_{x} N_{t}} \right),
    \\U_{x}^{I}(Q_{j}; t, x) &= \exp\left( 2 \pi i t \frac{Q_{j}}{N_{t}} \delta_{x, N_{x}} \right).
\end{aligned}
\end{equation}
The update procedure then assigns new values by setting
\begin{equation}
    U_{\mu}(t, x) \rightarrow U_{\mu}'(t, x) = U_{\mu}(t, x) U_{\mu}^{I}(Q_{j}; t, x) \quad \forall\, U_{\mu}(t, x) \in E(\Lambda),
\end{equation}
where $E(\Lambda)$ denotes the set of all link variables. Detailed balance can be achieved by adding a Metropolis accept-reject step; the inclusion of dynamical fermion would then necessitate the additional calculation of the determinant ratio of the new and old configuration. Finally, since this update procedure by itself is not ergodic, it must be combined with an ergodic update algorithm, e.g., a regular Metropolis link update.

\section{Multiscale thermalization}
Multiscale thermalization is a multigrid and real-space renormalization group (RG) inspired approach to accelerate the thermalization of configurations and to enable the generation of an ensemble of uncorrelated configurations more efficiently than through conventional approaches. Initially, configurations can be generated on a coarse lattice where the thermalization and autocorrelation times are shorter, before they are fine-grained onto a finer lattice. Since the fine-graining procedure is generally not an exact RG transformation, it is necessary to introduce rethermalization updates on the fine lattice. Even if the distribution of configurations is not exactly preserved by the fine-graining of the coarse ensemble, the rethermalization can in theory correct any deviations. Therefore the multiscale thermalization strategy is more efficient than conventional thermalization provided that the sum of the integrated autocorrelation time on the coarse lattice $\tau_{\mathrm{int}}^{c}$ and the rethermalization time $\tau_{r}$ is shorter than the integrated autocorrelation time on the fine lattice $\tau_{\mathrm{int}}^{f}$ (neglecting the initial thermalization times on both lattices).

Of course, this procedure is not limited to only two different lattice spacings, and the generalization to multiple levels with different lattice spacings is straightforward as long as the action can be matched between the different levels. Here, we only consider the simplest case of a single fine-graining transformation from a coarse lattice with spacing $2a$ to a fine lattice with spacing $a$. The fine-graining transformation used here attempts to interpolate the fields by first assigning values to those fine lattice links that can be matched to the coarse lattice links:
\begin{equation}
    U_{\mu}^{f}(\mathbf{n}) = U_{\mu}^{c}(\mathbf{n}/2), \quad U_{\mu}^{f}(\mathbf{n} + \hat{\mathbf{\mu}}) = \mathbbm{1} \qquad \forall\, \mathbf{n} \in V(\Lambda): \sum\limits_{\mu} (n_{\mu} \bmod 2) = 0.
\end{equation}
Here $\hat{\mu}$ denotes the orthonormal basis vector in $\mu$-direction, and $n_{\mu}$ is the $\mu$-component of the vector $\mathbf{n}$. Evidently, this transformation leaves all observables defined on the lattice corresponding to the coarse lattice invariant. Afterwards, the remaining links on the fine lattice are set to unity and smeared to reduce the occurring short range fluctuations of the field. Finally, a number of rethermalization updates are performed using conventional update algorithms.

\section{Results}
\begin{figure}[H]
    \centering
    \includegraphics[width=\textwidth]{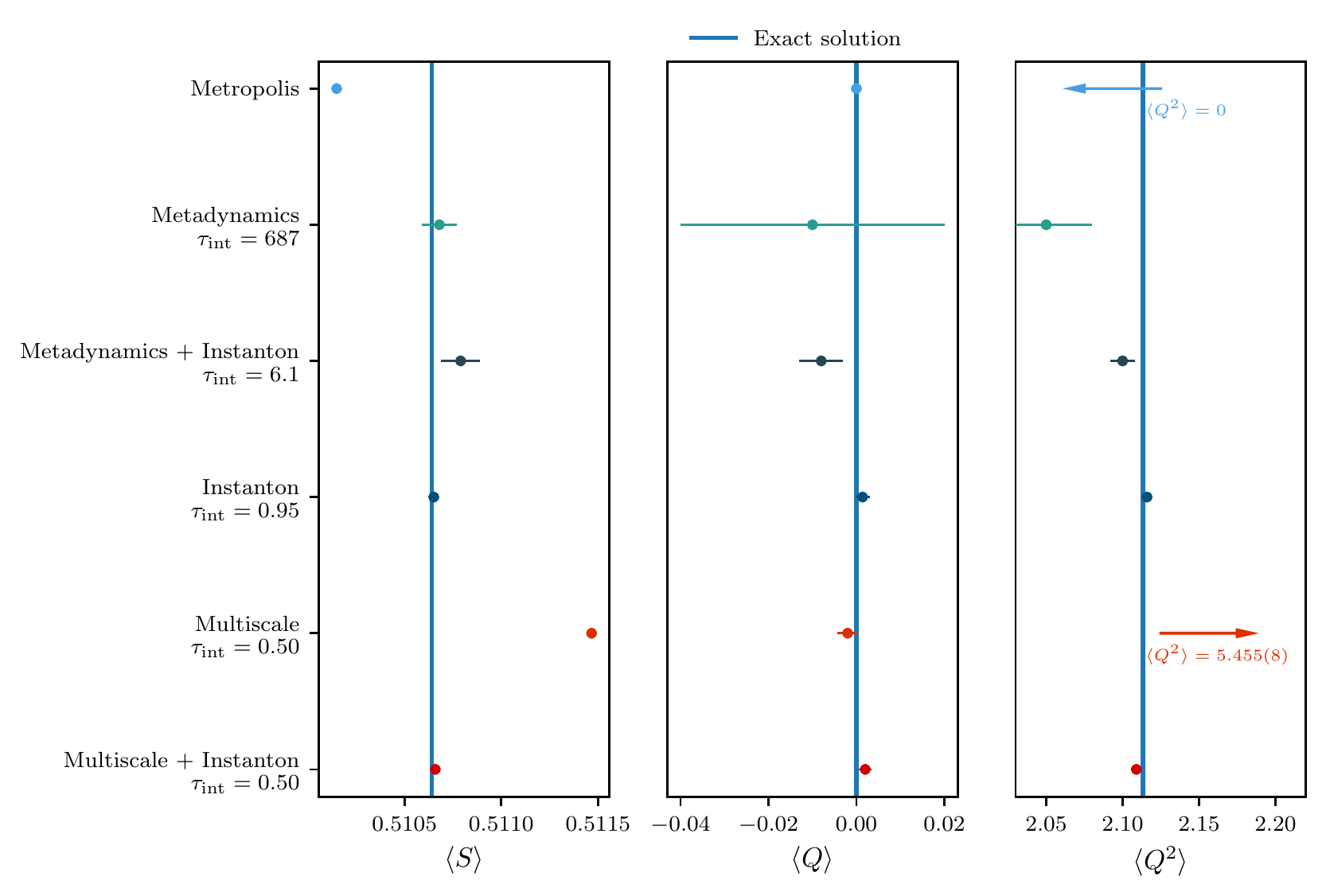}
    \caption{Comparison of expectation values obtained using the different approaches and combinations thereof. All simulations were performed on a $32^{2}$ lattice at $\beta = 12.8$, where conventional update algorithms are unable to sample different topological sectors properly. Additionally, analytical results \cite{Kovacs:1995nn, Elser:2001pe, Bonati:2019olo, Bonati:2019ylr} are indicated by the blue line.}
    \label{fig:final_comparison}
\end{figure}

\Cref{fig:final_comparison} and \cref{tab:final_comparison} show the expectation values of the action, the topological charge, and the topological susceptibility obtained using the different approaches and combinations thereof. With the exception of the two metadynamics approaches, all update schemes include overrelaxation updates. Additionally, the results are compared with analytical results \cite{Kovacs:1995nn, Elser:2001pe, Bonati:2019olo, Bonati:2019ylr}.

\begin{figure}[H]
    \centering
    \includegraphics[width=.7\textwidth]{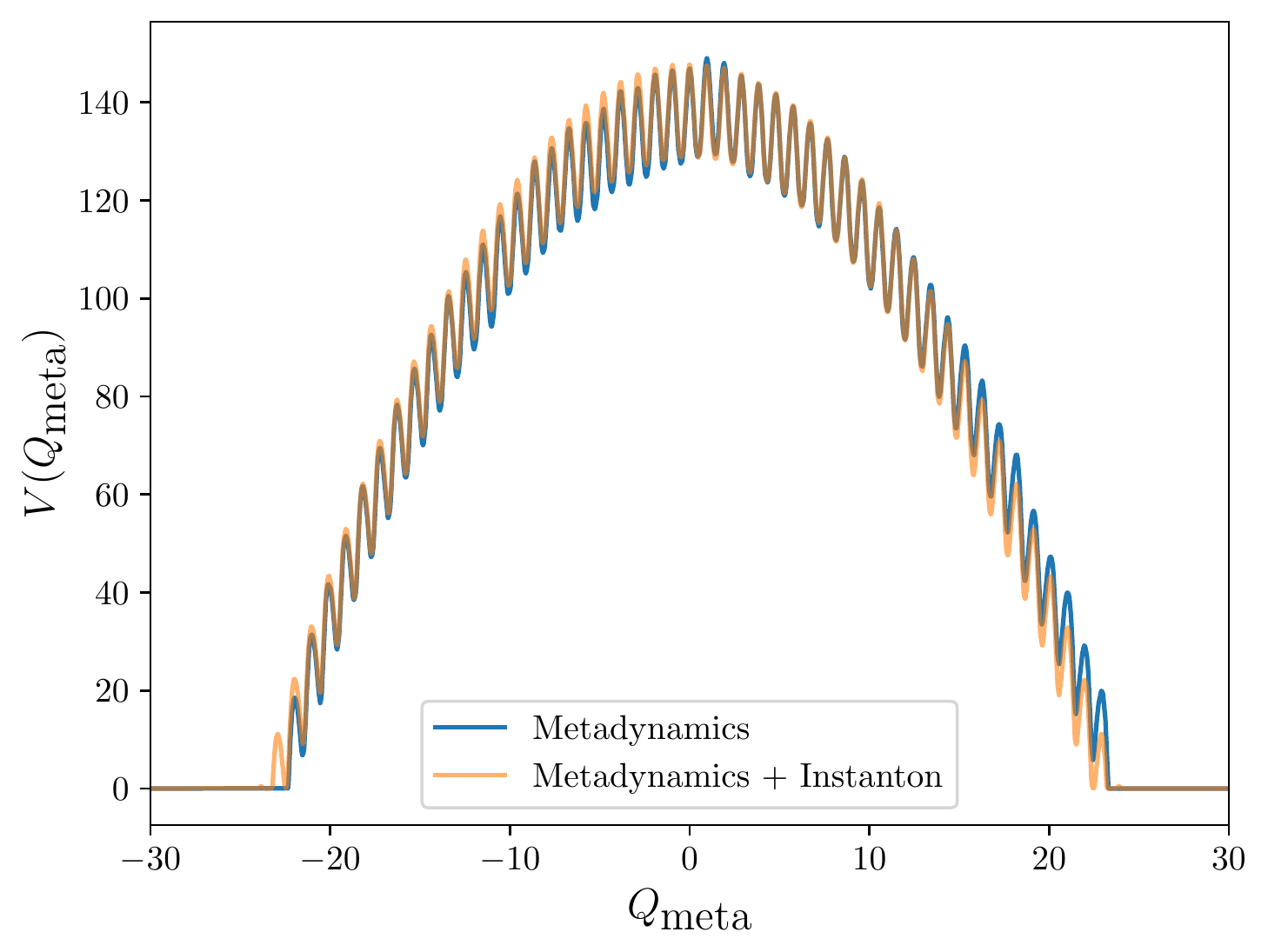}
    \caption{Comparison of the metapotentials in the range $[-25, 25)$ obtained using either multi-hit Metropolis updates or a combination of multi-hit Metropolis and instanton updates for a $32^{2}$ lattice at $\beta = 12.8$.}
    \label{fig:metapotential}
\end{figure}

Clearly, the standard Metropolis link update is unable to correctly sample different topological sectors at this lattice spacing, and the action as well as the topological susceptibility are underestimated, since all generated configurations were in the $Q = 0$ sector. On the other hand, all three approaches presented in the previous sections are able to unfreeze the topological charge and enable tunneling between the different topological sectors. However, both the metadynamics and the multiscale thermalization approach require several parameters to be tuned, whereas the instanton update can be applied directly. 
\begin{table}[H]
\centering
\begin{tabular}{ccccc}
	\bottomrule
	Approach & $\langle S \rangle$ & $\langle Q \rangle$ & $\langle Q^{2} \rangle$ & $\tau_{\mathrm{int}}$ \\ \bottomrule
	Analytical solution & $\num{0.510641}$ & $\num{0}$ & $\num{2.113304}$ & $-$ \\
	Metropolis & $\num{0.510149 \pm 0.000023}$ & $\num{0}$ & $\num{0}$ & $-$ \\
	Metadynamics & $\num{0.51068 \pm 0.00009}$ & $\num{-0.01 \pm 0.03}$ & $\num{2.05 \pm 0.03}$ & $687$ \\
	Metadynamics + Instanton & $\num{0.51079 \pm 0.00010}$ & $\num{-0.008 \pm 0.005}$ & $\num{2.100 \pm 0.008}$ & $6.1$ \\
	Instanton & $\num{0.510651 \pm 0.000023}$ & $\num{0.0014 \pm 0.0016}$ & $\num{2.116 \pm 0.003}$ & $0.95$ \\
	Multiscale & $\num{0.511467 \pm 0.000023}$ & $\num{-0.0020 \pm 0.0023}$ & $\num{5.425 \pm 0.008}$ & $0.50$ \\
	Multiscale + Instanton & $\num{0.510659 \pm 0.000022}$ & $\num{0.0020 \pm 0.0015}$ & $\num{2.109 \pm 0.003}$ & $0.50$ \\ \toprule
\end{tabular}
\caption{Comparison of expectation values obtained using the different approaches and combinations thereof, as well as an analytical solution. The results presented here are the same as in \cref{fig:final_comparison}.}
\label{tab:final_comparison}
\end{table}
We found that the best results with metadynamics were achieved by first building up the metapotential, and then simulating on a static potential in a separate run. This allows us to estimate the statistical errors in the usual way (for instance with block bootstrap resampling). The range of the metapotential defined by $Q_{\mathrm{thr,\, min}}$ and $Q_{\mathrm{thr,\, max}}$ requires some fine-tuning however. If the range is chosen too small, not all relevant topological sectors are properly sampled, but if the range is chosen too large, the number of effectively relevant configurations decreases, since the contribution of sectors with higher absolute values of topological charge to the expectation values are exponentially suppressed. Therefore, a lower bound for the range should be estimated by first simulating on a coarser, physically equivalent lattice where the topological sectors are still well-sampled. The metapotentials for a $32^{2}$ system at $\beta = 12.8$ obtained after \num{2e6} update sweeps using two different update schemes, namely multi-hit Metropolis updates with and without instanton updates, are shown in \cref{fig:metapotential}.
Both metapotentials clearly show the action barriers between topological sectors, as well as the overall increasing average action for larger absolute values of $Q_{\mathrm{meta}}$. Compared with using only multi-hit Metropolis updates, the potential obtained using combined multi-hit Metropolis and instanton updates is more symmetrical around the origin.\footnote{Of course, the metapotential can be symmetrized around $Q_{\mathrm{meta}} = 0$ \cite{Leinweber:2003sj} at the cost of the observable $\langle Q \rangle$.}

For the results in \cref{fig:final_comparison} and \cref{tab:final_comparison}, the metapotential parameters were chosen as follows:
\begin{equation}
    Q_{\mathrm{min}} = -10,\; Q_{\mathrm{max}} = 10,\; \delta Q = 0.001,\; w = 0.0001.
\end{equation}
The penalty potential parameters were set to the following values:
\begin{equation}
    k = 1000,\; Q_{\mathrm{thr,\, min}} = -8.0,\; Q_{\mathrm{thr,\, max}} = 8.0.
\end{equation}
The metapotentials were first generated using \num{2e6} update sweeps. Afterwards, the expectation values were calculated from \num{1e6} configurations separated by 10 update sweeps using a static metapotential (obtained using the final values of the metapotentials from the previous runs). All observables are compatible with the analytical results within $\sim 2$ standard deviations. However, the inclusion of instanton updates decreased the autocorrelation time by about two orders of magnitude and also reduced the statistical uncertainties of topological observables.

\begin{figure}[H]
    \centering
    \includegraphics[width=.7\textwidth]{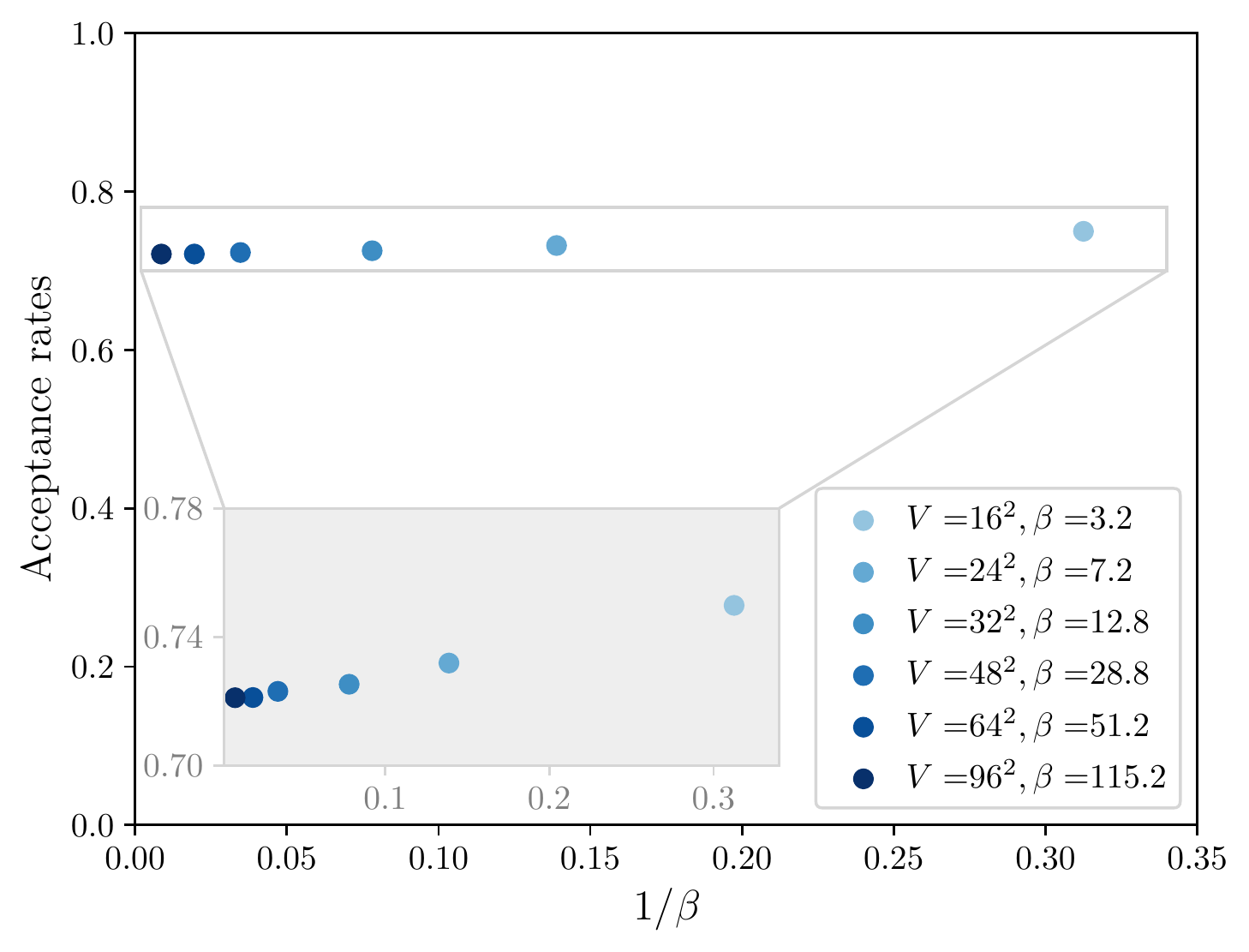}
    \caption{Acceptance rates of the instanton update in dependence of the gauge coupling. For all parameters examined here, the acceptance rate is nearly constant and lies between $0.72$ and $0.75$.}
    \label{fig:acceptance_rates}
\end{figure}

The application of the instanton update is straightforward, as no parameters have to be tuned. As shown in \cref{fig:acceptance_rates}, the acceptance rate is only weakly dependent on the lattice spacing and lies between $0.72$ and $0.75$ for all parameters examined here.
The expectation values shown in \cref{fig:final_comparison} and \cref{tab:final_comparison} were calculated from \num{1e6} configurations separated by 10 update sweeps. All expectation values are compatible with the analytical solution within one standard deviation, and consecutive configurations are nearly uncorrelated.

\begin{figure}[H]
    \centering
    \includegraphics[width=.495\textwidth]{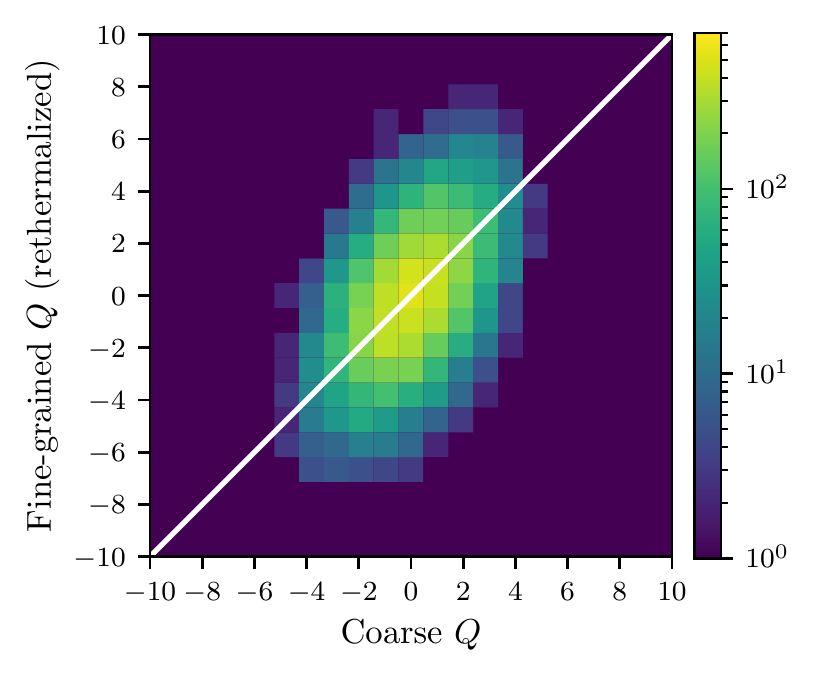}
    \includegraphics[width=.495\textwidth]{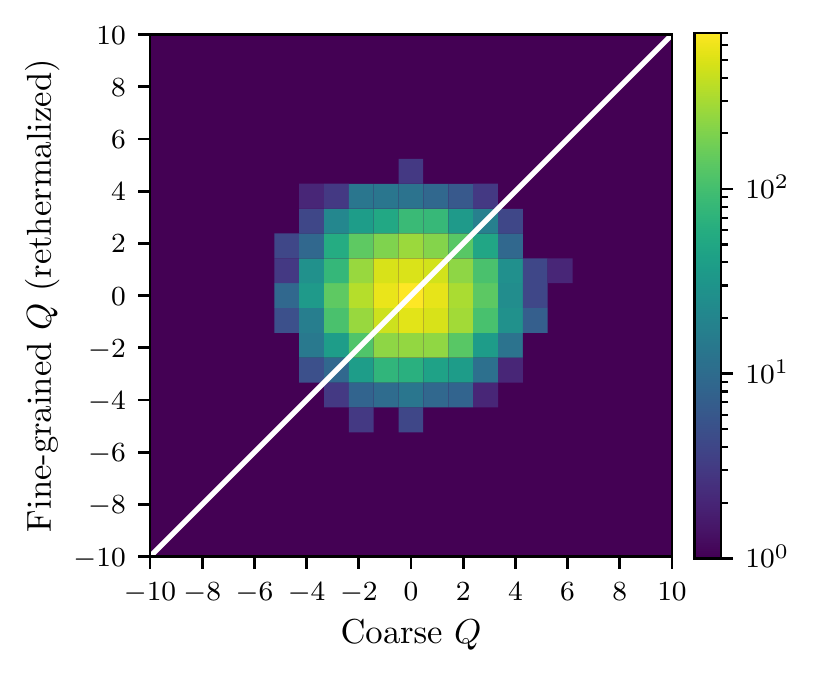}
    \caption{Charge correlation of configurations on the coarse lattice and configurations on the fine lattice after fine-graining and 100 rethermalization updates. The heatmap on the left side shows runs where the rethermalization was performed using multi-hit Metropolis and overrelaxation updates, whereas the heatmap on the right side shows runs where the rethermalization updates were performed using a combination of multi-hit Metropolis, overrelaxation, and instanton updates. The white lines are drawn to guide the eye.}
    \label{fig:charge_correlation}
\end{figure}

For the multiscale thermalization, we performed \num{1e8} updates on the coarse lattice. Every 100th configuration was fine-grained, and an additional 500 rethermalization updates were performed before measuring observables. This approach lead to correlation times slightly shorter than the instanton update, but in our case it did not sample the probability distribution of the fine lattice correctly. An important crosscheck is given by the correlation of the topological charge measured on the coarse lattice and on the fine lattice after fine-graining and rethermalization as seen in \cref{fig:charge_correlation}.
While the charges are clearly correlated, the charge is not exactly preserved, and instead spread out over a larger range on the fine lattice. In theory, this can be corrected by a longer rethermalization time, but since the topological charge is effectively frozen on the fine lattice, the required number of rethermalization updates makes this strategy unfeasible. The situation can be improved by including instanton updates. Despite the correlation between the two charges being less pronounced than before, the instanton update allows for tunneling between topological sectors and thereby corrects some of the deviations from the correct probability distribution during rethermalization. With this combined approach, all observables are compatible with the analytical solution within less than two standard deviations. However, it should be noted that runs with only 100 instead of 500 rethermalization updates overestimated the action by almost five standard deviations.

\section{Conclusion}
In these proceedings, we examined three algorithmic approaches towards alleviating the problem of topological freezing in the Schwinger model. The first approach - metadynamics - leads to higher autocorrelation times than the other approaches and requires some parameters to be tuned, but it allows one to control which parts of the configuration space are sampled. Instanton updates were found to produce the results most compatible with an analytical solution out of all update schemes studied here, and they do not require any parameters to be tuned. Finally, we found that the multiscale thermalization approach was unable to correctly sample the probability distribution of the fine lattice on its own for the parameters studied here. While that indicates this approach might not be suitable to properly sample all topological sectors at fine lattice spacings, it is still useful to reduce the thermalization and autocorrelation times of non-topological observables. We would also like to mention that a machine learning based approach to efficiently sample topological sectors in the Schwinger model has been presented at this conference \cite{Foreman:2021rhs}.

The extension of all three approaches to 4-dimensional SU(3) is straightforward, but does not always lead to similar levels of improvement as in the Schwinger model. Metadynamics or methods similar in spirit have been applied to high temperature QCD \cite{Bonati:2018blm} and pure SU(3) gauge theory \cite{Cossu:2021bgn}. The instanton update suffers from high action penalties, which effectively means that the acceptance is close to zero. Initial estimates show that the action penalty is of order $\sim 3000$ for a $8^{4}$ lattice at $\beta = 6.0$, and of order $\sim 12000$ for a $16^{4}$ lattice at $\beta = 6.0$ with Wilson gauge action. A possible remedy to this problem could be the use of gradient flow \cite{Luscher:2009eq}: By first flowing to positive flow time, applying the instanton update, and then flowing back in a symmetric way to respect detailed balance, the acceptance rates could possibly be enhanced. Finally, multiscale thermalization has already been successfully applied to pure SU(3) gauge theory and two-color QCD \cite{Endres:2015yca, Detmold:2016rnh}. Interestingly, recent studies suggest that rethermalization updates may not be necessary for very fine lattices \cite{Detmold:2018zgk}, which stands in contrast to our results in the Schwinger model.

\bibliographystyle{JHEP}
\bibliography{literature.bib}

\end{document}